\documentclass[twocolumn,english,aps,prb,twocolum,superscriptaddress,bibnotes,amsmath,amssymb,floatfix]{revtex4-1}
\usepackage[colorlinks=true,citecolor=blue,linkcolor=magenta]{hyperref}

\usepackage[markup=nocolor, authormarkupposition=left]{changes} 
\usepackage{soul}
\usepackage[utf8]{inputenc}
\usepackage[english]{babel}
\usepackage{amsmath,amsfonts,amssymb}
\usepackage[T1]{fontenc}
\usepackage{url}

\usepackage{amsmath}
\usepackage{amsfonts}
\usepackage{amssymb}

\usepackage{epstopdf}
\usepackage{graphicx}
\graphicspath{{./Figures/}}

\usepackage{changes}

\begin{document}
\title{Hybrid-integrated dark-pulse microcombs towards visible light spectrum}

\author{Jinbao Long}
\thanks{These authors contributed equally to this work.}
\affiliation{International Quantum Academy, Shenzhen 518048, China}

\author{Xiaoying Yan}
\thanks{These authors contributed equally to this work.}
\affiliation{International Quantum Academy, Shenzhen 518048, China}
\affiliation{College of Physics And Electronic Engineering, Shanxi University, Taiyuan 030006, China}

\author{Sanli Huang}
\thanks{These authors contributed equally to this work.}
\affiliation{International Quantum Academy, Shenzhen 518048, China}
\affiliation{Hefei National Laboratory, University of Science and Technology of China, Hefei 230088, China}

\author{Wei Sun}
\email[]{sunwei@iqasz.cn}
\affiliation{International Quantum Academy, Shenzhen 518048, China}

\author{Hao Tan}
\affiliation{International Quantum Academy, Shenzhen 518048, China}
\affiliation{Hefei National Laboratory, University of Science and Technology of China, Hefei 230088, China}

\author{Zeying Zhong}
\affiliation{International Quantum Academy, Shenzhen 518048, China}
\affiliation{Shenzhen Institute for Quantum Science and Engineering, Southern University of Science and Technology, Shenzhen 518055, China}

\author{Zhenyuan Shang}
\affiliation{International Quantum Academy, Shenzhen 518048, China}
\affiliation{Shenzhen Institute for Quantum Science and Engineering, Southern University of Science and Technology, Shenzhen 518055, China}

\author{Jiahao Sun}
\affiliation{International Quantum Academy, Shenzhen 518048, China}
\affiliation{Shenzhen Institute for Quantum Science and Engineering, Southern University of Science and Technology, Shenzhen 518055, China}

\author{Baoqi Shi}
\affiliation{International Quantum Academy, Shenzhen 518048, China}

\author{Chen Shen}
\affiliation{International Quantum Academy, Shenzhen 518048, China}
\affiliation{Qaleido Photonics, Shenzhen 518048, China}

\author{Yi-Han Luo}
\affiliation{International Quantum Academy, Shenzhen 518048, China}

\author{Junqiu Liu}
\email[]{liujq@iqasz.cn}
\affiliation{International Quantum Academy, Shenzhen 518048, China}
\affiliation{Hefei National Laboratory, University of Science and Technology of China, Hefei 230088, China}

\maketitle

\noindent\textbf{Leveraging hybrid integration, we demonstrate dark-pulse formation at 780-nm wavelength band in integrated Si$_3$N$_4$ microresonators driven by high-power AlGaAs-based chip-scale lasers. 
The device outputs coherent frequency combs with electronically detectable repetition rates down to 20 GHz, paving a route to efficient and compact atom-chip interfaces for spectroscopy, metrology and sensing.}

Microresonator-based frequency combs -- ``microcombs'' -- are miniaturized optical frequency combs (OFCs) with small size, weight and power consumption \cite{Kippenberg:18, Diddams:20}.
With the maturation of ultralow-loss integrated photonics, as well as hybrid and heterogeneous integration, today Kerr-nonlinear microresonators can be seamlessly integrated with high-power semiconductor lasers, high-speed photodetectors, and broadband modulators \cite{Gaeta:19, Liu:21, Stern:18, Xiang:21, Jin:21, Liu:20a, Sun:25}. 
Currently, microcombs have been predominantly studied and developed in the telecommunication band around 1550 nm, because of two facts. 
First, the technological maturity of integrated photonics culminates in the 1550-nm wavelength band for optical communication and datacenters. 
Second, many diode and fiber lasers, as well as EDFAs, are easily accessible, providing narrow-linewidth, high-power CW pumps for microcomb generation. 
As laser spectroscopy and frequency metrology with alkaline and alkaline-earth atoms are critical applications of OFCs, there is constant impetus to translate microcomb technology into the visible-light spectrum, e.g. around 780 nm for rubidium optical atomic clocks. 
Integrated microcombs could offer a viable and yet more convenient approach to comparing frequency uncertainty among different clocks. 

However, coherent (i.e. mode-locked) microcomb generation in the 780-nm wavelength band has not been equally fruitful. 
While Ref. \cite{Lee:17} has demonstrated a soliton microcomb driven by a 778-nm CW pump, the microresonator used is a suspended silica microdisk coupled by a tapered fiber, and thus is not integrated.
With Si$_3$N$_4$ integrated photonics, Ref. \cite{ZhaoY:20} has shown a 4-soliton-crystal microcomb. 
However, only 16 comb lines with $\sim4$ THz line spacing have been achieved, making frequency tuning for interrogating specific atomic transitions inconvenient. 
Similarly, Ref. \cite{Yu:19} has shown a single-soliton microcomb with 1 THz spacing, which is pumped at 1064 nm and emits a dispersive wave around 780 nm. 
Moreover, all these works have used many bulky fiber and free-space optics and lasers. 
None of these works has shown fully integrated microcombs with compact sizes.

\begin{figure*}[t!]
\centering
\includegraphics[width=0.9\linewidth]{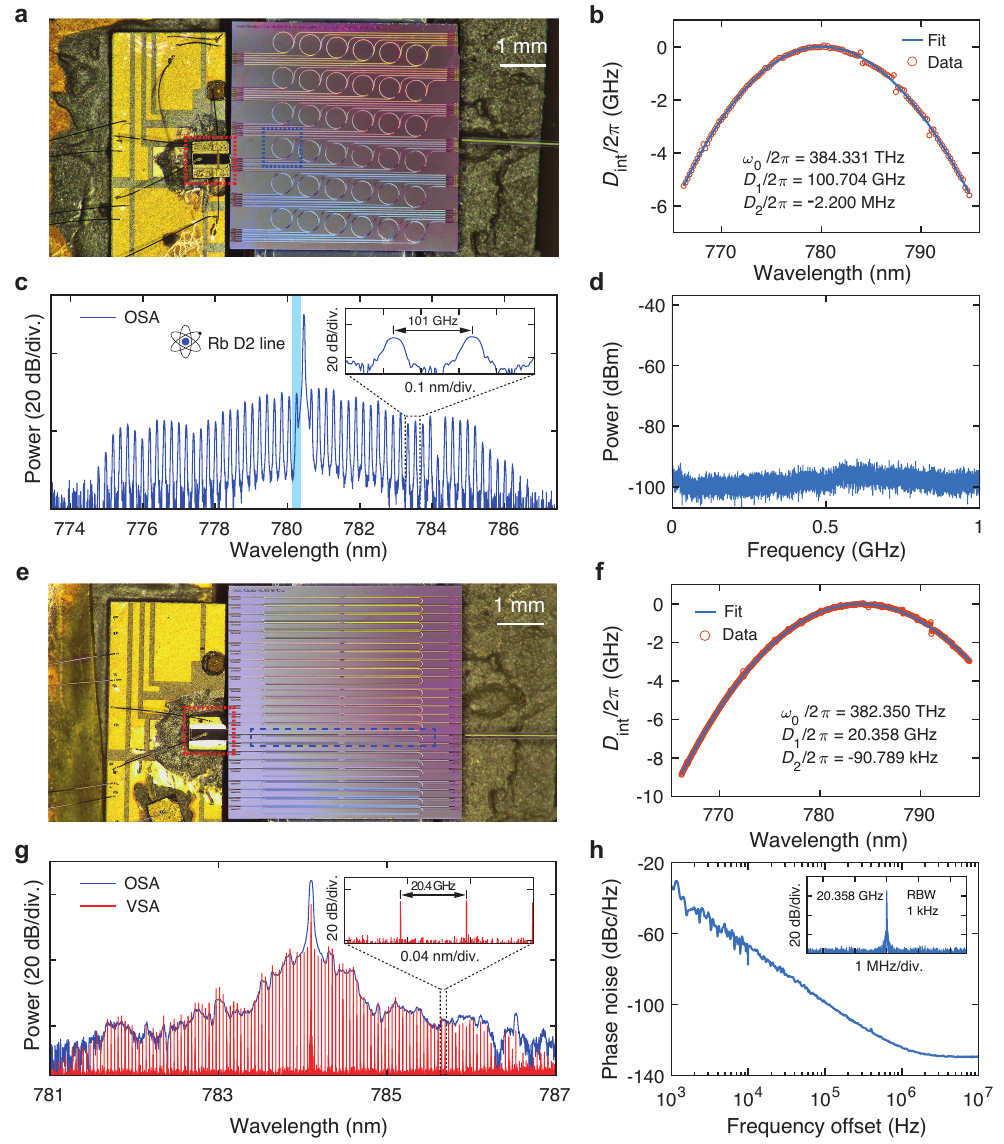}
\caption{
\footnotesize
\textbf{Dark-pulse microcomb formation in integrated microresonators driven by 780-nm CW lasers}.
\textbf{a, e}. 
Photograph showing that an AlGaAs-based laser diode (red rectangle) is edge-coupled to a Si$_3$N$_4$ chip containing dozens of microresonators with 100.7 (\textbf{a}) and 20.36 (\textbf{e}) GHz FSR. 
Light is coupled into only one of these microresonators (blue rectangle).
\textbf{b, f}.
Characterization of integrated dispersion of the Si$_3$N$_4$ microresonator in \textbf{a, e}.
The FSR is 100.7 (\textbf{b}) and 20.36 (\textbf{f}) GHz, and the GVD is normal.
\textbf{c}.
Optical spectrum of the dark pulse generated in \textbf{a} and recorded by an OSA. 
Inset highlights frequency spacing of $\sim101$ GHz. 
Blue-shaded zone marks the rubidium (Rb) D2 line. 
\textbf{d}.
Measured RF spectrum of the dark-pulse light in \textbf{c}.
The absence of amplitude noise indicates that the microcomb is a coherent dark-pulse train. 
\textbf{g}.
Optical spectrum of the dark pulse generated in \textbf{e}, which is recorded by the OSA (blue) and the VSA (red). 
Inset highlights frequency spacing of $\sim20.4$ GHz. 
\textbf{h}.
Measured phase noise of the 20.36-GHz microwave carrier converted from the microcomb's repetition rate in \textbf{g}. 
Inset shows the microwave's power spectrum with 1 kHz resolution bandwidth (RBW).
}
\label{Fig:1}
\end{figure*}

Here we demonstrate direct generation of coherent microcombs at 780 nm in ultralow-loss Si$_3$N$_4$ integrated microresonators. 
Figure~\ref{Fig:1}a shows the hybrid-integrated device where a high-power laser chip (red rectangle) is edge-coupled to a Si$_3$N$_4$ chip containing dozens of microresonators of 100 GHz FSR. 
The laser is a Fabry-P\'erot (FP)-type, AlGaAs-based laser diode emitting at 780 nm wavelength. 
The output laser power is 90 mW at 120 mA current. 
A printed circuit board (PCB) provides current and temperature stabilization to the diode. 

The Si$_3$N$_4$ chips are fabricated using a 150-mm-diameter (6-inch) foundry-level subtractive process of 300-nm-thick Si$_3$N$_4$.
Detailed fabrication process is found in Ref. \cite{Ye:23, Sun:25}. 
In particular, we installed a DUV stepper lithography (ASML PAS850C) with 110 nm resolution and used it to pattern the Si$_3$N$_4$ waveguides. 
The lithography resolution enables sufficiently small inverse tapers for efficient edge-coupling of 780-nm light from the laser to the Si$_3$N$_4$ chip \cite{Liu:18}. 
The measured coupling efficiency is 32\%, that is, with 4.9 dB loss per facet \cite{Shi:24}. 
The coupled Si$_3$N$_4$ microresonator (blue rectangle) is characterized by a visible-light vector spectrum analyzer (VSA) \cite{Shi:24}.
Figure~\ref{Fig:1}b shows the characterization result. 
The measured FSR is 100.704 GHz, and the group velocity dispersion (GVD) is normal / positive.

When the laser frequency is tuned into a resonance of the Si$_3$N$_4$ microresonator and light is coupled into the forward direction, Rayleigh scattering in the microresonator induces reflected light that is injected back to the laser. 
This ultrafast optical interaction triggers laser self-injection locking (SIL) \cite{Liang:15b, Kondratiev:23}, which passively and tightly locks the laser frequency to the Si$_3$N$_4$ microresonator resonance. 
When SIL occurs, microcombs can spontaneously form with sufficient intra-cavity optical power \cite{Kondratiev:23}. 
The normal GVD allows dark-pulse / platicon microcomb formation \cite{Xue:15, Lobanov:15}. 
Figure~\ref{Fig:1}c shows the dark-pulse spectrum recorded by a commercial optical spectrum analyzer (OSA, Yokogawa AQ6370D). 
The inset highlights the 100.7 GHz line spacing. 
Figure~\ref{Fig:1}d shows the RF spectrum of the microcomb light, detected by a low-noise photodetector and analyzed by an electronic spectrum analyzer. 
The absence of amplitude noise indicates that the microcomb is a coherent dark-pulse train \cite{Raja:19}. 

We further demonstrate SIL dark-pulse formation in a 20-GHz-FSR Si$_3$N$_4$ microresonator, as shown in Fig.~\ref{Fig:1}e. 
The characterized microresonator dispersion is shown in Fig.~\ref{Fig:1}f. 
The measured FSR is 20.358 GHz, and the GVD is again normal.
Figure.~\ref{Fig:1}g shows the dark-pulse spectrum recorded by both the commercial OSA and our VSA \cite{Shi:24}. 
The OSA fails to resolve the narrow-spaced comb lines, while the VSA succeeds due to its high frequency resolution (3 MHz).
The repetition rate of the microcomb is directly converted to a microwave carrier of 20.36 GHz via the low-noise photodetector.
The phase noise of the microwave carrier is analyzed and shown in Fig.~\ref{Fig:1}h with $-67$ dBc/Hz at 10 kHz Fourier frequency offset. 

In conclusion, we have demonstrated dark-pulse formation in integrated microresonators driven directly by 780-nm CW laser diodes. 
The microcombs feature 100.7 and 20.36 GHz repetition rates, which are more than an order of magnitude smaller than previous works \cite{ZhaoY:20, Yu:19} and are electronically detectable. 
Leveraging hybrid integration, our microcomb module occupies a footprint of only 29 mm$^2$.  
Meanwhile, both the Si$_3$N$_4$ microresonators and AlGaAs diodes can be manufactured in large-volume at low cost using standard CMOS\cite{Ye:23} and III-V foundries\cite{Yamashita:91}. 
Such cheap, compact, coherent microcombs operated in the visible-light wavelength can greatly facilitate ubiquitous deployment of integrated photonics for transportable atomic sensors and clocks on mobile platforms and in space applications.

\medskip
\begin{footnotesize}

\noindent \textbf{Funding Information}: 
We acknowledge support from the National Natural Science Foundation of China (Grant No. 12261131503, 12404436, 12404417), 
Innovation Program for Quantum Science and Technology (2023ZD0301500), 
National Key R\&D Program of China (Grant No. 2024YFA1409300),
Shenzhen Science and Technology Program (Grant No. RCJC20231211090042078), 
Shenzhen-Hong Kong Cooperation Zone for Technology and Innovation (HZQB-KCZYB2020050), 
and Guangdong-Hong Kong Technology Cooperation Funding Scheme (Grant No. 2024A0505040008).

\noindent \textbf{Acknowledgments}: 
We thank Yue Hu and Dengke Chen for assistance in chip characterization, 
and Zhen Chen for assistance in chip design. 

\noindent \textbf{Author contributions}: 
J. Long, X. Y., W. S. and B. S. built the setup, performed the experiment, and analyzed the data.
Y.-H. L. designed the Si$_3$N$_4$ chips. 
S. H., Z. Z., Z. S., J. S. and C. S. fabricated Si$_3$N$_4$ chips. 
H. T. provided the laser.
J. Long, W. S. and J. Liu wrote the manuscript, with the input from others.
J. Liu supervised the project.

\noindent \textbf{Disclosures} 
C. S. and J. Liu are co-founders of Qaleido Photonics, a start-up that is developing heterogeneous silicon nitride integrated photonics technologies. 
Others declare no conflicts of interest.

\noindent \textbf{Data Availability Statement}: 
The code and data used to produce the plots within this work will be released on the repository \texttt{Zenodo} upon publication of this preprint.

\end{footnotesize}
\bibliographystyle{apsrev4-1}
\bibliography{bibliography}
\end{document}